\documentclass[preprint,5p,twocolumn]{elsarticle}
\usepackage{amssymb}
\usepackage{amsmath}
\usepackage{mathrsfs}
\usepackage{natbib}

\usepackage{epsfig}
\usepackage{graphics}
\usepackage{float}
\usepackage{tabularx}
\usepackage{color}
\expandafter\ifx\csname package@font\endcsname\relax\else
 \expandafter\expandafter
 \expandafter\usepackage
 \expandafter\expandafter
 \expandafter{\csname package@font\endcsname}%
\fi
\newcolumntype{C}[1]{>{\centering\arraybackslash}p{#1}}

\begin{document}
 

\title{Class of Higgs-portal Dark Matter models in the light of gamma-ray excess from Galactic center}
\author[PRL]{Tanushree Basak}
\ead{tanu@prl.res.in}
\author[PRL,IIT]{Tanmoy Mondal}
\ead{tanmoym@prl.res.in}

\address[PRL]{Theoretical Physics Division, Physical Research Laboratory, Ahmedabad 380009, India.}
\address[IIT]{Department of Physics, Indian Institute of Technology, Gandhinagar, Ahmedabad, India.}

\def\be{\begin{equation}}
\def\ee{\end{equation}}
\def\al{\alpha}
\def\bea{\begin{eqnarray}}
\def\eea{\end{eqnarray}}
\def\beas{\begin{eqnarray*}}
\def\eeas{\end{eqnarray*}}

\begin{abstract}
Recently the study of anomalous gamma-ray emission in the regions surrounding the galactic center has drawn a lot of attention as it points out 
that the excess of $\sim 1-3$ GeV gamma-ray in the low latitude is consistent with the emission expected from 
annihilating dark matter. The best-fit to the gamma-ray spectrum corresponds to dark matter (DM) candidate  
having mass in the range $\sim 31-40$ GeV annihilating into $b\bar{b}$-pair 
with cross-section $\langle \sigma v \rangle = (1.4-2.0)\times 10^{-26}\;\textrm{cm}^3 \textrm{sec}^{-1}$. 
We have shown that the Higgs-portal dark matter models in presence of scalar resonance 
(in the annihilation channel) are well-suited for explaining these phenomena. In addition, the parameter space of these models 
also satisfy constraints from the LHC Higgs searches, relic abundance and direct detection experiments.  We also comment on real singlet scalar Higgs-portal DM model which is found to be incompatible with the recent analysis. 

\end{abstract}


\maketitle

\section{Introduction}

Gamma-ray emission from the galactic 
center (GC) and the inner galaxy regions as found in the Fermi-LAT data has gained a lot of attention from the perspective of 
dark matter (DM) searches. Past studies \cite{Goodenough:2009gk,Boyarsky:2010dr,Hooper:2010mq,
Hooper:2011ti,Abazajian:2012pn,Gordon:2013vta,Hooper_slatyer,Abazajian:2014fta} have pointed out a spatially extended 
excess of $\sim 1-3$ GeV gamma rays from the regions surrounding the galactic center, the morphology and spectrum 
of which is best fitted with that predicted from the annihilations of a $31-40$ GeV WIMP (weakly interacting massive particle)
dark matter (DM) candidate annihilating mostly to $b$-quarks  
(or a $\sim 7-10$ GeV WIMP annihilating significantly to $\tau$-leptons). Gamma rays from the galactic center is 
specially interesting because the region is predicted to contain very high densities of dark matter. Alternative explanations 
such as gamma-ray excess originating from thousands of unresolved millisecond pulsars have been disfavored since the 
signal extends well beyond the boundaries of the central stellar cluster. A more recent scrutiny of the morphology and 
spectrum of the anomalous gamma-ray emission in order to identify the origin has confirmed that the signal is very 
well fitted by a 31-40 GeV dark matter particle annihilating to $b\bar b$ with 
an annihilation cross section of $\sigma v = (1.4-2.0)\times 10^{-26} \textrm{cm}^3 \textrm{sec}^{-1}$ 
(normalized to a local dark matter density of $0.3$ GeV $\textrm{cm}^{-3}$) \cite{Daylan:2014rsa}, which is accidentally close to the 
weak cross-section for producing correct relic abundance. 

  The excess seen in the gamma ray spectrum at the low latitude region can be well explained in a simple dark matter model, where 
  the DM dominantly annihilates into quark pairs with cross-section in the desired range for obtaining correct relic abundance.
  Already a handful of particle physics model of dark matter \cite{daylan_ref3,daylan_ref4,daylan_ref1,daylan_ref2,
  daylan_cite1,daylan_cite2,daylan_cite3,daylan_cite4,daylan_cite5,daylan_cite6,daylan_cite7,daylan_cite8,park1,park2,
  Anchordoqui:2013pta,Izaguirre:2014vva,Abdullah:2014lla,Boehm:2014bia,Alves:2014yha,queiroz} have been proposed to explain the reported gamma-ray 
  excess. Among these some are focused on various Higgs-portal dark matter models \cite{daylan_ref4,daylan_ref1,daylan_cite7}. These kind of models are 
  simply interesting because they enjoy a special feature of scalar 
  resonances, provided dark matter mass is half of the scalar mass(es). This resonant feature is crucial as it 
   enhances the annihilation cross-section. 
  
  In this letter, we have studied a class of Higgs-portal dark matter models to explain the reported excess.
 We showed that the simplest Higgs-portal model, i.e, the real singlet scalar extension of the 
  Standard model (SM), is inconsistent with a $30-40$ GeV dark matter, because of the absence of resonance. Another Higgs-portal model considered in this letter is the so-called Singlet fermionic dark matter (SFDM) model, which 
  consist of SM alongwith a hidden sector with a gauge singlet scalar and a Dirac-fermion singlet, acting as a potential 
  DM candidate.  We analyse the parameter space of this model owing to constraints from LHC bound on SM-Higgs, relic density and direct 
  detection of DM. We found this model to be consistent as well with the requirements to explain the galactic center $\gamma$-ray 
  excess.   The last model 
  we consider is the minimal $U(1)_{B-L}$ extension of the SM with a SM singlet scalar $S$ 
  and three right-handed (RH) neutrinos. The third generation RH-neutrino, which is a 
  Majorana fermion, serves as  a viable DM candidate as an artifact of $\mathbb{Z}_2$-symmetry. The parameters like DM coupling with the SM-Higgs boson and scalar mixing are subject to the constraints from the LHC Higgs searches apart from  other observational constraints on dark matter. However, annihilation of Majorana fermionic dark matter through a scalar resonance is velocity suppressed. But, the presence of a very narrow scalar resonance in the DM annihilation channel lifts the cross sections considerably
  via Breit-Wigner enhancement at later times and makes the model compatible with the recent analysis.

\section{Class of Higgs-portal dark matter models}
\label{sec:higgs-portal}
The basic feature of Higgs-portal model is that all the interactions of DM are mediated through Higgs(es) and the presence of 
scalar resonance plays a crucial role in determining the correct relic abundance. Here, we will discuss a class of Higgs-portal 
DM model in the light of the recent analysis  \cite{Daylan:2014rsa} of the excess gamma-ray emission in the Fermi-bubble.

\subsection{Scalar Singlet extension of SM}
\label{sec:singlet_dm}

The scalar singlet extension of SM \cite{McDonald:1993ex,Burgess:2000yq,Davoudiasl:2004be,Bandyopadhyay:2010cc,Guo:2010hq,
He:2011gc,Cline:2013gha} is the most simplified Higgs-portal model to account for a WIMP candidate. The real singlet 
$S^\prime$, stabilized by odd $\mathbb{Z}_2$-parity, acts as a viable DM candidate. It interacts only with the SM Higgs boson through the renormalizable interaction term present in the lagrangian,
\begin{equation}
 \mathcal{L}=\mathcal{L}_{SM}+ \frac{1}{2}(\partial {S^\prime})^2- \frac{1}{2}\mu^2_{S^\prime} {S^\prime}^2+ 
 \mathcal{L}_{int}- \lambda {S^\prime}^4
\end{equation}
where, $ \mathcal{L}_{int}= -\lambda_{S^\prime}\lvert \Phi \rvert^2 {S^\prime}^2$.

\begin{figure}[t!]
 \centering
 \includegraphics[
 keepaspectratio=true,scale=0.65]{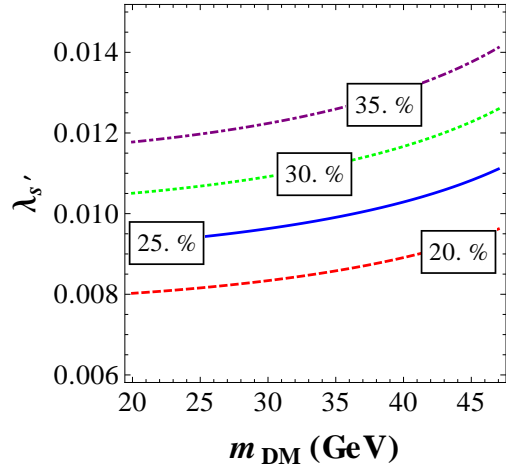}
 \caption{Contours of invisible branching ratio for singlet scalar DM model, in the plane of $\lambda_{S^\prime} - m_{DM}$. 
 Region above red-dashed (blue-solid, green-dotted and purple-dot-dashed) line is excluded if the SM Higgs has 
 invisible branching ratio upto 20\% (25\%, 30\% and 35\%.).}
 \label{fig:singlet_dm}
\end{figure}

The mass of the DM after EWSB becomes, $m^2_{DM}=\mu^2_{S^\prime}+ \frac{1}{2} \lambda_{S^\prime}v^2$. 
The coupling between DM 
and SM-Higgs, i.e, $\lambda_{S^\prime}$ is constrained from the invisible decay width of Higgs boson when  
$m_{S^\prime}\lesssim m_h /2$, such that BR$(h \to SS)\lesssim 0.20$ \cite{higgs-inv-decay}. Figure.~\ref{fig:singlet_dm} shows 
the contours of invisible branching ratio of the SM Higgs boson in $\lambda_{S^\prime} - m_{DM}$ plane. Region 
above red-dashed  line is excluded as in the region the invisible branching ratio of the SM Higgs is more than 20\%. 
Blue-solid, green-dotted and purple-dot-dashed contours show the allowed region if the invisible branching ratio is 25\%, 30\% and 35\% 
respectively. As expected, the more invisible decay, the higher values of $\lambda_{S^\prime}$ are allowed.
For example,  $\lambda_{S^\prime}$ must be $\lesssim 8\times 10^{-3}$ if 20\% of the SM Higgs decays invisibly .

\subsubsection{Relic Abundance}
 The relic abundance of DM can be formulated as \cite{Kolb_Turner},  
\be\label{eq:relic_density}
\Omega_{_{CDM}} h^2 = 1.1\times 10^9 \frac{x_f}{\sqrt{g^*}m_{Pl}\langle\sigma v\rangle_{ann}} \textrm{GeV}^{-1}\; , 
\ee
where $x_f =m_{DM}/T_D$ with $T_D$ as decoupling temperature. $m_{Pl}$ is Planck mass  =  $1.22\times 10^{19}$ GeV, and, $g^*$ is effective
number of relativistic degrees of freedom. $\langle\sigma v\rangle_{ann}$ is the thermal averaged value of DM annihilation
cross-section times relative velocity. 
$\langle\sigma v\rangle_{ann}$ can be obtained using the well known formula \cite{Srednicki:1988ce}, 
\be\label{eq:thermal_avg}
\langle\sigma v\rangle_{ann} = \frac{1}{m_{DM}^2}\bigg\{ w(s) - 
\frac{3}{2}\Big(2 w(s) - 4{m_{DM}^2} w'(s)\Big) \frac{1}{x_f} \bigg\}, 
\ee
where prime denotes differentiation with respect to $s$ ($\sqrt{s}$ is the center of mass energy) and 
evaluated at $s=\big(2m_{DM}\big)^2$. The function $w(s)$ is same as defined in \cite{tanmoy}.

In order to fit the spectrum of the gamma-ray emission near the galactic center, one requires a WIMP of mass 
$\sim 31-40$ GeV, which dominantly annihilates into final state $b\bar{b}$ through the s-channel exchange of the SM-Higgs boson
. Also we choose, $ \lambda_{S^\prime}\simeq 0.007$ as a benchmark value.  We obtain that 
$ \langle \sigma v\rangle_{b\bar{b}}= (0.92-2.17)\times 10^{-30}\;\textrm{cm}^3/s$ 
, which cannot fit the observed gamma-ray signal. We also found that such a WIMP candidate cannot produce the required 
relic-abundance unless a scalar resonance is present  
i.e, when, $m_{S^\prime}\simeq m_h /2 \sim 62$ GeV. Also Ref.\cite{He:2011gc} has mentioned that for $m_{S^\prime}< m_h /2$, the parameter space is severely restricted from both LHC and direct detection constraints. 
 We conclude that 
the singlet scalar DM with mass around 31-40 GeV is incompatible with the dark matter interpretation for the
gamma ray excess from GC. 


\subsection{Singlet fermionic dark matter model}
\label{sec:sfdm}

The singlet fermionic dark matter (SFDM) model is a renormalizable extension of SM with a hidden sector containing a scalar singlet 
$\Phi_s$ and a singlet Dirac fermion $\psi$ \cite{Kim:2008pp,Baek:2012uj}. Here, the singlet fermionic dark matter 
$\psi$, interacts with the SM sector via the singlet $\Phi_s$ which mixes with the SM-Higgs doublet $\Phi$. Therefore, 
this is also an example of Higgs-portal model. The lagrangian 
of the SFDM model is given as,
\begin{equation}
 \mathcal{L} = \mathcal{L}_{SM} + \mathcal{L}_{hid} + \mathcal{L}_{int} 
\end{equation}
where,
\begin{eqnarray}
 \mathcal{L}_{hid} &=& \mathcal{L}_{\Phi_s}+\bar{\psi}(i\partial_\mu \gamma^\mu -m_\psi)\psi -\lambda_{\psi S}\; \bar{\psi} \psi \Phi_s \\
 \mathcal{L}_{int} &=& \frac{\lambda_1^\prime}{2}\Phi^\dagger \Phi \Phi_s +\frac{\lambda_{2}^\prime }{2}\Phi^\dagger \Phi \Phi_s^2 \\ 
 \mathcal{L}_{\Phi_s} &=&  \frac{1}{2}(\partial \Phi_s)^2 -\frac{m_{\Phi_s}^2}{2}\Phi_s^2 
 -\frac{\lambda^\prime}{3}\Phi_s^3- \frac{\lambda''}{4}\Phi_s^4 
\end{eqnarray}

After EWSB, the singlet field  $\Phi_s$ can be written as, $\Phi_s=x+s$, where $x$ is the VEV of $\Phi_s$ and $\Phi=(0 \;\;v+\phi)^T$.  
 The two scalar eigenstates are denoted as,
\begin{eqnarray}
 H_2 &=& \sin\alpha \; s + \cos\alpha \; \phi \\
 H_1 &=& \sin\alpha \; \phi - \cos\alpha \; s 
\end{eqnarray}
where, $H_2$ is identified as the SM-Higgs boson and we consider the case when, $m_{H_2}>m_{H_1}$. Now, the mass of the DM is given by, 
$m_{DM}=m_\psi+ \lambda_{\psi S}\; x$, with $m_\psi$ as a free parameter. In order to explain the observed gamma-ray excess in the low latitude, we consider the 
following set of parameters, $m_{DM} \sim 31$ GeV, $m_{H_1}\simeq 2m_{DM}$.
The DM interaction strength depends on the 
parameter $\lambda_{DM}=\lambda_{\psi S}$. Thus, the two parameters $\lambda_{DM}$ and scalar mixing $\cos\alpha$ play 
crucial role in DM phenomenology. Here, the scalar mixing angle and DM-coupling  are subject to various constraints 
like LHC bound on SM-Higgs boson, relic abundance  of DM and upper bound on the DM-nucleon scattering cross section.

\subsubsection{Constraints from LHC}

\begin{figure}[t!]
 \centering
 \includegraphics[
 keepaspectratio=true,angle=0,scale=0.65]{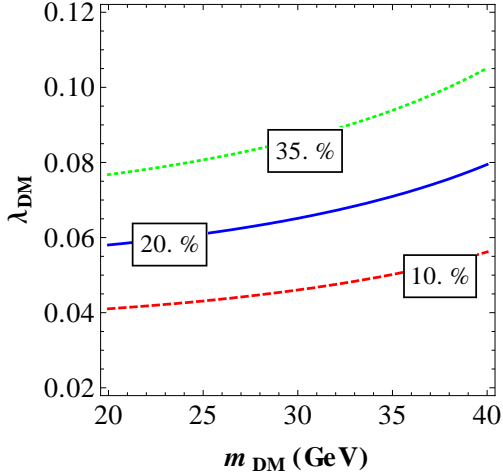}
 \caption{ Contours of invisible branching ratio for SFDM model (10\%, 20\%, and 35\%) in the plane of 
 [$m_{DM}, \lambda_{DM}$] with $\cos\alpha=0.95$. }
 \label{fig:SFDM-invbr-r2}
\end{figure}

Observation of SM-like Higgs boson at LHC by CMS \cite{cmshiggs} and ATLAS \cite{atlashiggs} collaboration 
will constrain this mixing angle severely. 
The signal strength or reduction factor of a particular channel can be defined as: 
\be \label{eq:ratio}
r_i^{xx}  = \frac{\sigma_{H_i}}{\sigma_{H_i} ^{SM}}\cdot \frac{BR_{H_i \to xx}}{BR_{H_i \to xx} ^{SM}}\;, \; (i=1,2).
\ee 
where, $\sigma_{H_i}$ and $BR_{H_i \to xx}$ are the production cross section of $H_i$ , and the branching ratio of
$H_i \to xx$ respectively. Similarly, $\sigma_{H_i} ^{SM}$ and $BR_{H_i \to xx} ^{SM}$ are the corresponding quantities
of the SM-Higgs. Using eq.~\ref{eq:ratio} one obtains, 
\begin{eqnarray}
\label{eq:r1-r2}
r_2 &=& \cos^4\alpha \frac{\Gamma_{H_2}^{SM}}{\cos^2\alpha\; \Gamma_{H_2}^{SM} + \sin^2\alpha\; \Gamma_{H_2}^{Hid} + 
\Gamma_{H_2 \to H_1 H_1}}\nonumber \\
r_1 &=& \sin^4\alpha \frac{\Gamma_{H_1}^{SM}}{\sin^2\alpha\; \Gamma_{H_1}^{SM}+ \cos^2\alpha\; \Gamma_{H_1}^{Hid}}
\end{eqnarray}
where, $\Gamma_{H_i}^{SM} $ denotes the total decay width of the SM-Higgs boson and $\Gamma_{H_i}^{Hid} $ is the 
invisible decay width ($H_i \to 2$ DM). 
The invisible decay width of the SM Higgs reads as
\be\label{eq:H2inv}
\Gamma_{H_2}^{Hid}\equiv \Gamma_{inv} = \frac{m_{H_2} \,\lambda_{DM}^2}{16\pi} \sin^2\alpha \left(1-4\frac{m_{DM}^2}{m_{H_2}^2} \right)^{\frac{3}{2}},
\ee

Since, $m_{DM}< m_{H_2}/2$, we can constrain the DM coupling $\lambda_{DM}$ from the invisible decay width of 
SM-Higgs boson. Figure.\ref{fig:SFDM-invbr-r2} shows the allowed range of $\lambda_{DM}$ with mass of DM 
for different invisible branching ratio of the SM-Higgs boson, assuming the width of the Higgs to
SM fermions as 4.21 MeV. We observe that for $m_{DM}\sim 30$ GeV, if 
$BR_{inv}\geq 20$\% (35\%) then DM-coupling, $\lambda_{DM}$ should be less than 0.06 (0.075). Again, the signal 
strength (as defined in eqs. \ref{eq:ratio}-\ref{eq:r1-r2}) 
depends on the scalar mixing angle. Constraining 
$r_2$ to be $\leq 0.9$ (or 0.8), we obtain the allowed range of scalar mixing $\cos\alpha$ as a function of $m_{DM}$ 
for a particular value of DM-coupling.

\subsubsection{Constraints from relic density and direct detection}
\label{sec:sfdm-relic}

We obtain the relic abundance (using Eqn.~\ref{eq:relic_density}) of the dark matter in agreement with WMAP-9 year result 
\cite{wmap9} and PLANCK \cite{planck}, only near resonance where, 
$m_{DM}=m_{h_1}/2\sim 31$ GeV. Dominant contribution to relic density comes from final-state $b\bar b$ annihilation with cross-section 
$\langle \sigma v \rangle \simeq 1.7 \times 10^{-26} \textrm{cm}^3\textrm{sec}^{-1}$, which is also in the desired range for 
explaining galactic center $\gamma$-ray excess. 
We observe that as we decrease $\lambda_{DM}$, the annihilation cross-section is also decreased. But, if we approach 
very near the resonance region, i.e, $m_{H_1}-2m_{DM}\sim \mathcal{O}(10^{-4})$, 
the annihilation cross-section can be enhanced significantly, which counter-balance the previous effect. 
However, if we are slightly away from resonance we need to have $\lambda_{DM}\sim 10^{-2}$, to get correct relic. 

\begin{figure}[t!]
 \centering
 \includegraphics[
 keepaspectratio=true,angle=0,scale=0.59]{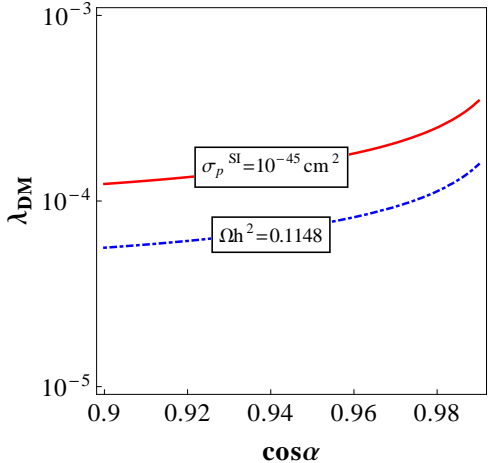}
  \caption{ Contours of relic abundance (blue-dot-dashed) consistent with WMAP9 \cite{wmap9} and 
  spin-independent scattering cross-sections for SFDM model, $\sigma_p^{SI}=10^{-45} \textrm{cm}^2 \; (\textrm{red-solid})$
,  in the plane of 
 [$\lambda_{DM}, \cos\alpha$] for $m_{DM}\sim 31$ GeV. }
 \label{fig:SFDM-relic-sigma}
\end{figure}

The scattering cross-section (spin-independent) for the dark matter off a proton or neutron as,
\begin{equation}
\label{eq:sigmaSI}
 \sigma_{p,n}^{SI}=\frac{4m_r^2}{\pi}f_{p,n}^2
\end{equation}
where, $m_r$ is the reduced mass defined as, $1/m_r=1/m_{DM}+1/m_{p,n}$ and  $f_{p,n}$ is the hadronic matrix element, given by
\begin{equation}
\label{scalarterms}
f_{p,n} = \sum_{q=u,d,s}  f_{Tq}^{(p,n)} a_q  \frac{m_{p,n}}{m_q}  + \frac{2}{27}f_{TG}^{(p,n)}
\sum_{q=c,b,t} a_q \frac{m_{p,n}}{m_q} \nonumber.
\end{equation}
The f-values are given in \cite{Ellis:2000ds}. Here, $a_q$ is the effective coupling constant between the DM and the quark. 
An approximate form of $a_q/m_q$ can be recast as :
\begin{eqnarray}
\label{eq:aqmq}
 \frac{a_q}{m_q} &=& \frac{\lambda_{DM}}{v\sqrt{2}} \Bigg[\frac{1}{m_{h}^2}-\frac{1}{m_{H}^2}\Bigg]
 \sin\!\alpha \cos\!\alpha \; .
\end{eqnarray}
 In order to be consistent with the latest exclusion limit on $\sigma_p^{SI}$ as specified by LUX \cite{lux}, {\sc Xenon 100} \cite{Lavina:2013zxa,Aprile:2012nq}, 
we require $\sigma_p^{SI} \lesssim 10^{-45} \textrm{cm}^2$. In Figure.~\ref{fig:SFDM-relic-sigma}, we show the contour of $\sigma_p^{SI}=10^{-45} \textrm{cm}^2 \; (\textrm{red-solid})$. 
It indicates that $\lambda_{DM}$ should be small 
 enough (in the range of $\sim 10^{-4} - 10^{-5}$) to satisfy the required value of $\sigma_p^{SI}$. 
 As argued before, very near resonance region, for $\lambda_{DM}\sim 10^{-4} $, also gives correct relic density. The contour of relic abundance has been shown in Figure.~\ref{fig:SFDM-relic-sigma} by the blue-dot-dashed line.
 
\subsection{Minimal $U(1)_{B-L}$ gauge extension of SM}
\label{sec:b-l-dm}

The minimal $U(1)_{B-L}$ extension of the SM \cite{Khalil,Basso:2010jm,Basso:2011hn,tanmoy} contains in addition to SM : a SM singlet $S$ with $B-L$ charge +2, three right-handed 
neutrinos $N_R^i (i=1,2,3)$ having $B-L$ charge -1. The assignment of $\mathbb{Z}_2$-odd charge ensures the stability of $N_R^3$ \cite{Okada:2010wd,Basak} which qualified as a viable DM candidate.
 Scalar Lagrangian of this model can be written as, 
\begin{equation}\label{eq:new-scalar_L}
\mathcal{L}_s=\left( D^{\mu} \Phi\right) ^{\dagger} D_{\mu}\Phi + \left( D^{\mu} S\right) ^{\dagger} D_{\mu}S - V(\Phi,S ) \, ,
\end{equation}
where the potential term is, 
\be\label{BL-potential}
V(\Phi,S) = 
m^2\Phi^{\dagger}\Phi + \mu ^2|S|^2 + \lambda _1 (\Phi^{\dagger}\Phi)^2 +\lambda _2|S|^4 + \lambda _3 \Phi^{\dagger}\Phi|S| ^2  \nonumber \, ,
\ee 
with $\Phi$ and $S$ as the SM-scalar doublet and singlet fields, respectively. After spontaneous symmetry breaking (SSB) the  singlet scalar field can be written as, $S=\frac{v_{_{B-L}}+\phi'}{\sqrt{2}}$ with $v_{_{B-L}}$ real and positive. 
The mass eigenstates ($H_1,H_2$) are linear combinations of $\phi$ and $\phi^{\prime}$ with mixing angle $\alpha$.  
 We identify $H_2$ as the SM-like Higgs boson with mass 125.5 GeV. We choose $v_{_{B-L}}\simeq 4$ TeV, in 
accordance with the constraint on the mass of $Z'$-boson \cite{pdg2012}.

The scalar mixing angle, $\alpha$ can be expressed as:
\be \label{eq:theta}
\tan(2\alpha) = \frac{\lambda_3 v_{_{B-L}} v}{\lambda_1 v^2 - \lambda_2 v_{_{B-L}}^2}.
\ee

The RH neutrinos interact with the singlet scalar field $S$ through interaction term of the lagrangian: 
\begin{eqnarray}
\label{yuk}
 \mathcal{L}_{int} &=& \sum_{i=1}^3\frac{y_{n_i}}{2}\overline{N_R^i} S N_R^i.
\end{eqnarray}
Here we define, $\lambda_{DM}$ as the coupling between DM candidate and the SM Higgs boson, which 
 is effectively the Yukawa coupling of the $N_R^3$. 
Thus, the mass of dark matter is given by, $m_{DM}=m_{N_R^3}=\frac{y_{n_3}}{\sqrt 2}v_{_{B-L}}$.

\subsubsection{Constraints from LHC}

As $\lambda_{DM}$ is suppressed by $B-L$ symmetry breaking VEV, the invisible decay width remains very small ($\sim 0.5\%$) 
for DM mass $\sim 30-40$ GeV. 

On the other hand, the decay width of the SM Higgs decays to light scalar boson is 
\be\label{eq:H2H1H1}
\Gamma_{H_2\to H_1\,H_1} = \frac{g_{_{H_2H_1H_1}}^2 }{32\pi\,m_{H_2}} \sqrt{1-4\frac{m_{H_1}^2}{m_{H_2}^2}}.
\ee
where $g_{_{H_2H_1H_1}}$ is defined in \cite{tanmoy}. 
In order to have $H_2$ as a SM Higgs boson we require $r_2 \geq 0.9$ (0.8)
and correspondingly $r_1 \leq 0.1$ (0.2).  We have obtained that $r_2$ being $\geq 0.9$ (0.8) restricts the choice of 
scalar mixing such that, $\cos\alpha \geq 0.96$ (0.94) for $m_{DM}\sim 31$ GeV.


\begin{figure}[t!]
 \centering
 \includegraphics[
 keepaspectratio=true,angle=-90,scale=0.35]{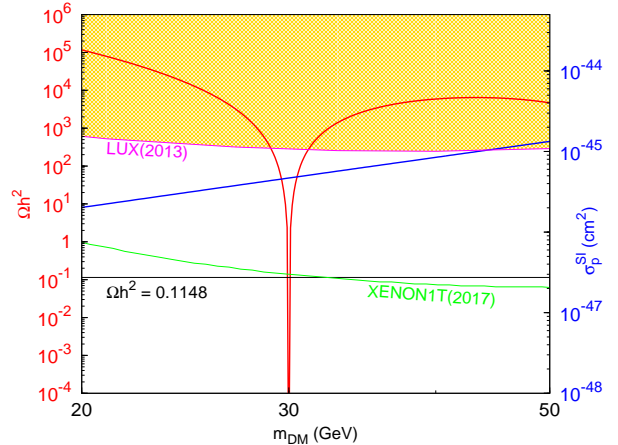}
 \caption{Shows the relic abundance (red curve) and scattering cross-section (blue curve) as a function of DM mass. The black (solid) line shows the 
 latest 9-year WMAP data i.e, $\Omega_{_{CDM}} h^2 = 0.1148\pm 0.0019$ \cite{wmap9}. The yellow region above is excluded by LUX(2013) 
 \cite{lux} and the green (dashed) line shows the projected sensitivity of {\sc Xenon1T} experiment \cite{xenon1T}.
 }
 \label{fig:b-l-relic-SI}
\end{figure}


\subsubsection{Velocity dependent cross-section and Breit-Wigner enhancement}

In general the annihilation of Majorana fermionic DM into SM-fermion pairs through a scalar mediator is velocity 
suppressed. In that case the thermally averaged annihilation cross-section can be written as,
\begin{equation*}
 \langle \sigma v\rangle = a +bv^2 \; , \textrm{where $a,b$ are model dependent variables.}
\end{equation*}
The term $a$ comes from s-channel s-wave process, where as, $b$ has contributions from
both s-wave and p-wave. The averaged velocity $v$ can be expressed as, $v\sim \sqrt{3/x}$. Because of p-wave suppression, $\langle \sigma v\rangle$ at the time of freeze-out ($x_f\sim 20$) is different than that at the 
galactic halo ($x\sim 10^6$). However, $\langle \sigma v\rangle$ at the galactic halo can be substantially enhanced using the Breit-Wigner mechanism \cite{murayama,guo}, where the DM annihilates through a narrow s-channel resonance. 

The leading annihilation channels of DM are, 
${N_{R}^3} {N_{R}^3} \longrightarrow b\bar{b},\,\tau^+ \tau^-$.
The s-channel resonant annihilation cross-section into final state $b\bar{b}$ (dominant) is given as,
\begin{eqnarray*}
 4E_1E_2\sigma v &=& \frac{1}{8\pi}\sqrt{1-\frac{4m_b^2}{s}}|\bar{M}|^2 \nonumber \\
                 &=& \frac{\lambda_{DM}^2\cos^2\alpha}{32\pi^2}\frac{s^2}{m_{H_1}^2}\frac{m_{H_1}\Gamma_{H_1}}{(s-m_{H_1}^2)^2+m_{H_1}^2\Gamma_{H_1}^2}
\end{eqnarray*}
where, $\Gamma_{H_1}$ is the total decay width of $H_1$.

Here, we introduce two parameters $\delta$ and $\gamma$ as,
\begin{equation}
m_{H_1}^2=4m_{DM}^2(1-\delta)\; , \; \gamma=\Gamma_{H_1}/m_{H_1}.
\end{equation}
Clearly, $\delta<0$ and $\delta>0$ represents the physical and unphysical pole respectively. 
Adopting the single-integral formula for thermally averaged cross-section, we obtain,
\begin{eqnarray}
\label{eq:bw}
 \langle \sigma v\rangle &=& \frac{1}{n_{EQ}^2}\frac{m_{DM}}{64\pi^4x}\int_{4m_{DM}^2}^\infty ds\;4E_1E_2\sigma v \sqrt{s}g_i^2 
  \nonumber \\ 
 && \times\;\sqrt{1-\frac{4m_{DM}^2}{s}}\;K_1\!\left(\frac{x\sqrt{s}}{m_{DM}}\right)
\end{eqnarray}
where,\begin{equation*}
 n_{EQ}=\frac{g_i}{2\pi^2}\frac{m_{DM}^3}{x}K_2(x) 
\end{equation*}
$K_1(x)$ and $K_2(x)$ are the modified Bessel's function of second kind and $g_i$ is the internal degrees of freedom of dark matter
particle.

We again redefine $s$ as, $s=4m_{DM}^2(1+y)$ where, $y\propto v^2$. 
Eq.\ref{eq:bw} can be recast in terms of $\delta$, $\gamma$ and $y$ as,
\begin{equation}
 \langle \sigma v\rangle\propto x^{3/2}\int_0^{y_{eff}}\frac{\sqrt{y}(1+y)^{3/2}e^{-xy}}{(y+\delta)^2+\gamma^2(1-\delta^2)}dy
\end{equation}
where, $y_{eff}\sim \textrm{max}[4/x, 2|\delta|]$ for $\delta<0$ and $y_{eff}\sim 4/x$ for $\delta>0$ case. 
If $\delta$ and $\gamma$ are much smaller than unity, $\langle \sigma v\rangle$ scales as $v^{-4}$ in the
limit $v^2\gg\textrm{max}[\gamma,\delta] $. At smaller velocity, the thermally averaged 
annihilation cross-section becomes proportional to $v^{-2}$ and approach towards a constant value when 
$v^2\ll \textrm{max}[\gamma,\delta] $.  

We obtain the relic abundance using eqs.(\ref{eq:relic_density},\ref{eq:bw}). 
Fig.\ref{fig:b-l-relic-SI} shows the relic abundance (red curve) as a function of 
DM mass. The resultant relic abundance is found to be consistent with the reported value of WMAP-9 \cite{wmap9} (shown by 
the black solid line) and PLANCK experiment \cite{planck},  only near resonance when, $m_{DM} \sim (1/2)\; m_{H_1}$.

We have also achieved the required $\langle \sigma v\rangle_{b\bar{b}} \sim 1.881\times10^-26\; \textrm{cm}^3/s$ at the galactic halo 
through the Breit-Wigner enhancement given the value of parameters\footnote{Also for positive values of $\delta$ (for example, $\delta\simeq 10^{-1}$ and $\gamma \simeq 10^{-5}$), it is possible to obtain the required boost factor \cite{murayama,guo}} $\delta\simeq -10^{-3}$ and $\gamma \simeq 10^{-5}$. Note that, 
the same set of parameter values have been used to compute the relic abundance.

\subsubsection{Constraints from direct detection searches}

The spin-independent scattering cross-section of DM off nucleon is obtained using eq.\ref{eq:sigmaSI}. In Fig.\ref{fig:b-l-relic-SI} 
 the yellow region above is excluded by LUX(2013) \cite{lux}. 
We observe that the resultant spin-independent scattering cross-section (blue curve) lies well below the LUX exclusion limit. However, the projected sensitivity of {\sc Xenon1T} experiment \cite{xenon1T} (green-dashed line) 
might constrain the scenario of $m_{DM}=31-40$ GeV in this model.


\section{Summary and Conclusion}
\label{sec:summary}

The excess of $\gamma$-ray emission in the low latitude region near the 
galactic center can be explained by annihilation of DM (in the mass range $\sim 31-40$ GeV) into $b\bar b$, with cross-section 
of the order of the weak cross-section (i.e $\sim 10^{-26}\textrm{cm}^3\textrm{sec}^{-1}$). In this context, we have analysed a class of Higgs-portal 
DM models and constrain the parameter space of these models.  We found that the real singlet scalar DM model is incompatible with 
the recent analysis. However, the singlet fermionic dark matter model can account for this 
phenomena apart from satisfying relic abundance criterion. Besides this, the SI-scattering cross-section can be well below the 
exclusion limit from LUX, {\sc Xenon 100}, provided $\lambda_{DM}$ lies below $\sim 10^{-4}$. 
Also, RH-neutrino DM in the minimal $U(1)_{B-L}$ model is well-suited for explaining the galactic-center gamma-ray excess along with satisfying other DM and collider constraints. The relic abundance is
found to be consistent with the recent WMAP9 and PLANCK data only near scalar resonances, i.e, 
$m_{DM} \simeq m_{H_1}/2$.  Here, 
we obtain the required $\langle \sigma v \rangle$ for explaining this reported excess at the galactic center through Breit-Wigner enhancement mechanism. Although, future experiment like {\sc Xenon 1T} can further restrict the parameter space of minimal 
$U(1)_{B-L}$ model.

In passing by, we would like to mention that the anti-proton data from indirect detection experiments like PAMELA \cite{pamela,Cirelli-pamela}, AMS-02 \cite{ams02} have constrained the annihilation cross-section into hadronic (mostly $b\bar b$) final states in a model independently way. But, the 
present exclusion limit on $\langle\sigma v\rangle_{b\bar b}$ lies much above the reported value in Ref.\cite{Daylan:2014rsa}  for DM mass 
in the range 31-40 GeV. However, the bound on $\langle\sigma v\rangle_{b\bar b}$ from the projected anti-proton data of AMS-02 
(see Fig.1 of Ref.\cite{park1}) can be an important discriminator of dark matter models.

\section*{Acknowledgements}

We would like to thank Partha Konar and Subhendra Mohanty for most useful comments and discussions. 

\bibliographystyle{apsrev}
\bibliography{Ref}

\end{document}